\documentclass{ws-procs975x65}

\usepackage{graphicx}
\usepackage{euscript,amssymb}
\usepackage{epsfig}
\usepackage{color}
\usepackage{amsfonts}
\usepackage{amsmath}
\usepackage{amssymb}
\usepackage{fancyhdr}
\usepackage{esint}

\usepackage{colordvi}

\begin{document}

\title{LOOP QUANTUM COSMOLOGY FOR NONMINIMALLY COUPLED SCALAR FIELD}

\author{MICHA{\L} ARTYMOWSKI$^*$, ANDREA DAPOR$^\dagger$ and TOMASZ PAW{\L}OWSKI$^\ddagger$}

\address{$^*$ Instytut Fizyki Teoretycznej, Uniwersytet Warszawski, ul. Ho\.{z}a 69, 00-681 Warsaw, Poland\\
E-mail: michal.artymowski@fuw.edu.pl\\
$^\dagger$ Instytut Fizyki Teoretycznej, Uniwersytet Warszawski, ul. Ho\.{z}a 69, 00-681 Warsaw, Poland\\
E-mail: adapor@fuw.edu.pl\\
$^\ddagger$ Katedra Metod Matematycznych Fizyki, Uniwersytet Warszawski, ul. Ho\.{z}a 74, 00-682 Warsaw, Poland\\
E-mail: tomasz.pawlowski@fuw.edu.pl}

\begin{abstract}
We perform a LQC-quantization of the FRW cosmological model with nonminimally coupled scalar field. Making use of a canonical transformation, we recast the theory in the minimally coupled form (Einstein frame), for which standard LQC techniques can be applied to find the physical Hilbert space and the dynamics. We then focus on the semiclassical sector, obtaining a classical effective Hamiltonian, which can be used to study the dynamics. We show that the classical singularity is replaced by a "mexican hat"-shaped bounce, joining the contracting and expanding branches. The model accommodates Higgs-driven inflation, with more than enough e-folding for any physically meaningful initial condition.
\end{abstract}


\bodymatter
\section{Classical model and transformation to Einstein frame}
 {We consider the theory of gravity coupled to a scalar field as specified by action}
\begin{eqnarray} \label{action}
S[\phi, g_{\mu \nu}] = \dfrac{1}{8 \pi G} \int d^4x \sqrt{-g} \left[-U(\phi) R + \dfrac{1}{2} \partial_\mu \phi \partial^\mu \phi - V(\phi)\right],
\end{eqnarray}
where the non-minimal coupling is realized through $U(\phi) = (1 + \xi \phi^2)/2$, with $\xi$ being a coupling constant\cite{1}. 
 {Our point of focus is} the FRW sector of the theory (\ref{action}): the canonical variables parametrizing this part of the phase space are the scale factor $a$, the homogeneous scalar field $\phi$, and the associated momenta $\pi_a$ and $\pi_\phi$.
\\
We perform the canonical transformation $(g_{\mu \nu}, \phi) \rightarrow (\tilde{g}_{\mu \nu}, \tilde{\phi})$ such that
\begin{eqnarray} \label{E-frame}
  \tilde{g}_{\mu \nu} = 2U g_{\mu \nu}, \ \ \ 
  \left(\dfrac{d\tilde{\phi}}{d\phi}\right)^2 = \dfrac{U + 3U'^2}{2U^2};\ \ \ 
   {V\to\tilde{V}=\frac{V(\phi(\tilde{\phi})}{4U^2(\phi(\tilde{\phi})}}.
\end{eqnarray}
Under this change,  {$S[g_{\mu \nu}, \phi]$ transforms to an action $\tilde{S}[\tilde{g}_{\mu \nu}, \tilde{\phi}]$ 
where $\tilde{\phi}$ (with a modified potential $\tilde{V}$) is \emph{minimally} coupled to gravity $\tilde{g}_{\mu \nu}$.}
\section{LQC effective theory}
With respect to the tilded variables -- which  {correspond to} the so-called Einstein frame -- the theory reduces to a scalar field minimally coupled to gravity. The loop quantization of the cosmological sector of such theory is well known\cite{2}. We first define Ashtekar-Barbero variables  {adapted to isotropic setting, which in turn are respresented by a canonical pair}:
\begin{eqnarray} \label{AB-vars}
  \text{sgn}(\tilde{v}) \tilde{v} = \dfrac{\tilde{a}^3}{2\pi \gamma \sqrt{\Delta} \ell_{\text{Pl}}^2}, 
  \ \ \ \
  \tilde{b} = - \gamma \sqrt{\dfrac{\Delta}{2U}} \dfrac{\dot{\tilde{a}}}{\tilde{a}},
  \ \ \ \
   {\{\tilde{b},\tilde{v}\} = -2/\hbar,}
\end{eqnarray}
where $\Delta = 4 \sqrt{3} \pi \gamma \ell_\text{Pl}^2$ and $\gamma$ is the Barbero-Immirzi parameter. The choice of quantizing these Einstein variables instead of the old ones, $(v, b)$, is not only driven by the experience in dealing with minimally coupled systems, but it is also suggested on a more fundamental level. Indeed, Einstein frame preserves all symmetries of general relativity (and in particular local Lorentz symmety); the hypothesis is that the non-minimal description is an ''emergent'' one, which should arise from the underlying theory following the quantization in the Einstein frame.
\\
The LQC quantization of this model prescribes a kinematical Hilbert space of the form $\mathcal{H}_\text{kin} = \mathcal{H}_\text{gr} \otimes \mathcal{H}_\phi$, where $\mathcal{H}_\text{gr} = L^2(\overline{\mathbb{R}}, d\mu_\text{Bohr})$ and $\mathcal{H}_\phi = L^2(\mathbb{R}, d\phi)$. A  {convenient} basis for $\mathcal{H}_\text{gr}$ is provided by the eigenstates $| \tilde{v} \rangle$ of the volume operator $\widehat{\tilde{V}} := \widehat{\tilde{a}}^3$, defined as
\begin{eqnarray} \label{q-vol}
  \widehat{\tilde{V}} | \tilde{v} \rangle 
  = 2\pi \gamma \sqrt{\Delta} \ell_\text{Pl}^2 \widehat{\tilde{v}} | \tilde{v} \rangle 
  = \alpha  {\tilde{v}} | \tilde{v} \rangle.
\end{eqnarray}
The scalar constraint is represented as an operator on $\mathcal{H}_\text{kin}$, given by $\hat{\bm{H}} = \hat{\bm{H}}_\text{gr} \otimes \mathbb{I}_\phi + \hat{\bm{H}}_\phi$ where
\begin{eqnarray} \label{q-ham}
\hat{\bm{H}}_\text{gr} = \dfrac{3\pi G}{8\alpha} \sqrt{|\hat{\tilde{v}}|} \left(\hat{\tilde{N}}^2 - \hat{\tilde{N}}^{-2}\right)^2 \sqrt{|\hat{\tilde{v}}|}, \ \ \ \ \ \hat{\bm{H}}_\phi = \dfrac{1}{2\alpha} \widehat{|\tilde{v}|^{-1}} \pi_{\tilde{\phi}}^2 + \dfrac{\alpha}{\hbar} |\hat{\tilde{v}}| \dfrac{V}{4U^2},
\end{eqnarray}
where  $\hat{\tilde{N}}$  {is the shift operator}  {on $\mathcal{H}_\text{gr}$ such that} $\hat{\tilde{N}} | \tilde{v} \rangle = | \tilde{v} + 1\rangle$.
\\
Upon deparametrization with respect to irrotational dust\cite{3}, this constraint acting on a kinematical Hilbert space is turned into a true Hamiltonian acting on  {$\mathcal{H}_\text{gr} \otimes \mathcal{H}_\phi$ which now becomes} the physical Hilbert space of the theory. The dynamics is then defined by the  {Schr\"odinger} equation
\begin{eqnarray} \label{q-dyn}
-i\hbar \partial_{\tilde{t}} \Psi(\tilde{v}, \tilde{\phi}) = \hat{\bm{H}} \Psi(\tilde{v}, \tilde{\phi}).
\end{eqnarray}
At this point we have at our disposal a complete quantum theory for the system. However, since the aim of our work is to study the corrections to the classical picture, we consider the so-called \emph{effective dynamics}\cite{4}. This consists heuristically into the replacement of operators $\hat{\tilde{v}}$ and $\hat{\tilde{N}}$ with their expectation values on a semiclassical state. The result is the ''classical'' effective Hamiltonian
\begin{eqnarray} \label{e-ham}
\bm{H}_\text{eff} = -\dfrac{3\pi G}{8\alpha} |\tilde{v}| \sin^2(\tilde{b}) + \dfrac{\pi^2_{\tilde{\phi}}}{2\alpha |\tilde{v}|} + \dfrac{\alpha |\tilde{v}|}{\hbar} \dfrac{V}{4U^2} = E_\text{eff},
\end{eqnarray}
where $E_\text{eff}$ is the energy of the dust clock field.
\section{Dynamics}
Having an effective Hamiltonian, we can solve Hamilton equations for the system.  {For technical reasons it is easier 
to evolve the ``mixed frame'' set of variables $(\tilde{v}, \tilde{b}, \phi, \pi_{\tilde{\phi}})$ and then extract the data corresponding to a particular frame (tilded/untilded).} 
 {The initial data are set at the (always present) bounce point $\tilde{b} = \pi/2$ in Einstein frame, which point also provides the time origin $t=0$. Because the equations are homogeneous in $\tilde{v}$, we can freely set its initial value to $1$. Finally, the remaining variable $\pi_{\tilde{\phi}}$ is fixed via (\ref{e-ham}) with $E_\text{eff}=0$ to eliminate the dynamic influence of the 
dust clock field. As a consequence all the solutions are labeled by one free data: $\phi_\text{in} := \phi(t = 0)$.} 
 {The resulting initial value problem (equations + initial data) is then solved numerically. The principal results are presented on 
figure~1. They are: $(i)$ the universe evolution consists of the expending and contracting classical phases joined deterministically by the epoch of modified evolution featuring two consecutive bounces and a recollapse (with scale factor following a ``mexican hat'' curve); $(ii)$ the number $N$ of e-foldings during inflation is more than sufficient to conform with observations} ($N \gtrsim 60$) for a wide range of values, confirming the classical behaviour of the model even after quantum gravity effects are taken into account.
\begin{figure}
\begin{centering}
\includegraphics[height=1.6in]{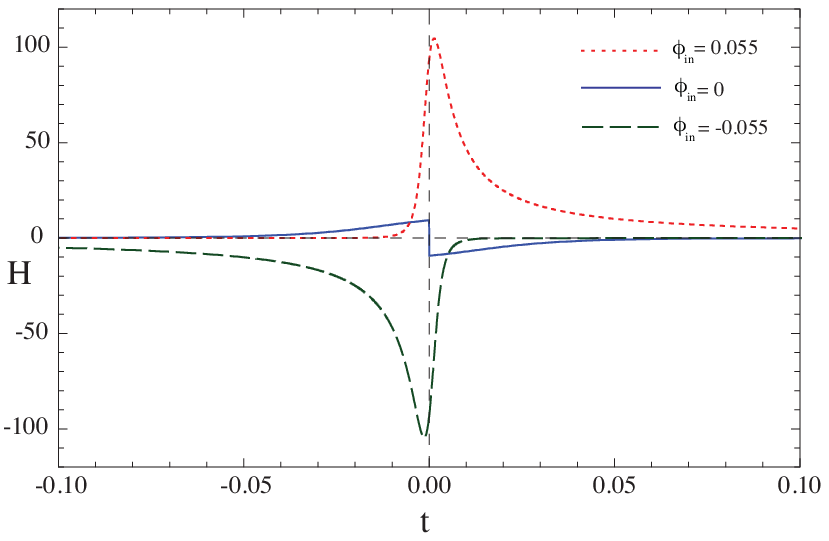} \includegraphics[height=1.6in]{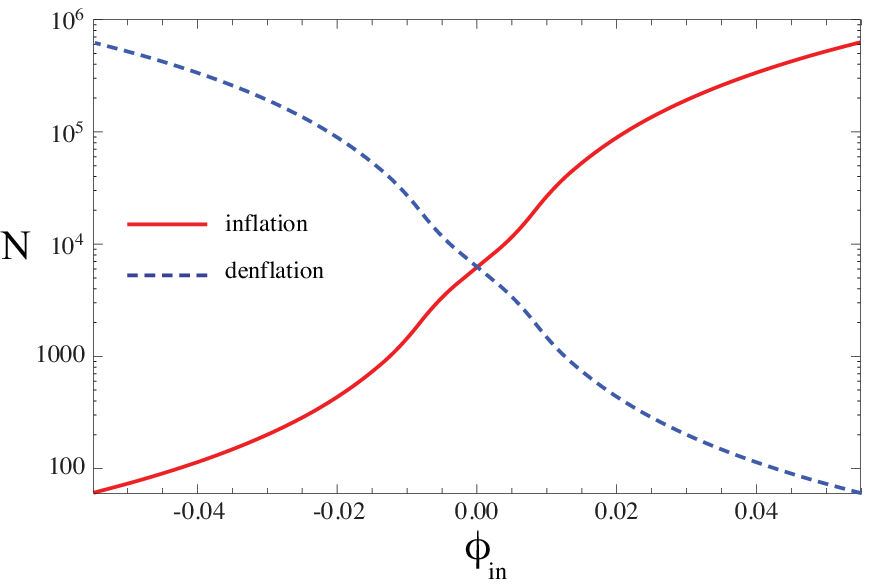}
\caption{{\footnotesize (1a) Evolution of the Hubble parameter as a function of time (in Planck units) for the symmetric trajectory ($\phi_{\text{in}} = 0$) and two non-symmetric ones. (1b) Number of e-foldings during the slow-roll inflation after the bounces (red line) and the deflation before the bounces (blue line) as a function of initial value of the field (labelling the solutions). Notice that even the solutions with very high time reversal asymmetry (considered to be unlikely) give more than $60$ e-foldings during inflation.}}
\par\end{centering}
\end{figure}
%
\section*{Acknowledgments}
%
The work was supported in parts by grants of Minister Nauki i Szkolnictwa Wy{\.z}szego no.~182/N-QGG/2008/0, 
N202~091839 and N202~104838 and by National Science Centre of Poland grant DEC-2011/01/M/ST2/02466.

\end{document}